\documentclass[pra,onecolumn,floatfix,superscriptaddress,longbibliography,notitlepage, nofootinbib]{revtex4-1}
\pdfoutput=1

\usepackage[utf8]{inputenc}
\usepackage{braket}
\usepackage{amsmath}
\DeclareMathOperator{\Tr}{Tr}
\usepackage{dcolumn}   % needed for some tables
\usepackage{bm}        % for math
\usepackage{amssymb}
\usepackage{mathtools}
\usepackage{tikz}
\usepackage{qcircuit}
\usepackage{appendix}
\usepackage{hyperref}
\usepackage{subcaption}
\usepackage{amsmath,amsfonts,amssymb}
\usepackage{mathtools}
\usepackage{thmtools,thm-restate}
\usepackage{algorithm}

\usepackage{algpseudocode}

\newtheorem{theorem}{Theorem}
\newtheorem{lemma}{Lemma}

\usetikzlibrary{arrows.meta}

\def\BibTeX{{\rm B\kern-.05em{\sc i\kern-.025em b}\kern-.08em
    T\kern-.1667em\lower.7ex\hbox{E}\kern-.125emX}}

\begin{document}
\title{ Alternative Method for Estimating Betti Numbers}
\author{Nhat A. Nghiem}
\affiliation{Department of Physics and Astronomy, State University of New York at Stony Brook, Stony Brook, NY 11794-3800, USA}
\affiliation{C. N. Yang Institute for Theoretical Physics, State University of New York at Stony Brook, Stony Brook, NY 11794-3840, USA}

\begin{abstract}
    Topological data analysis (TDA) is a fast-growing field that utilizes advanced tools from topology to analyze large-scale data. A central problem in topological data analysis is estimating the so-called Betti numbers of the underlying simplicial complex. While the difficulty of this problem has been established as NP-hard, previous works have showcased appealing quantum speedup. In this article, we provide an alternative method for estimating Betti numbers and normalized Betti numbers of given simplicial complex, based on some recent advances in quantum algorithm, specifically, quantum singular value transformation. Our method can be faster than the best-known classical method for finding Betti numbers, and interestingly, it can also find the Betti numbers of the complement graph to our original one. Comparing to the best known quantum algorithm, our method generally requires lower depth circuit, in trade-off for longer running time. Regarding normalized Betti numbers, our method could match the running time of best-known quantum method in the case of dense simplices.
\end{abstract}
\maketitle

\section{Introduction}
The realm of quantum computation has developed quickly since its initial development~\cite{shor1999polynomial, grover1996fast, deutsch1985quantum, deutsch1992rapid}, resulting in a diverse and far-reaching impact on multiple disciplines. For example, the quest for fault-tolerance quantum computation has fostered both theoretical and experimental efforts in realizing and manipulating anyons~\cite{kitaev1995quantum}. Besides, there is a surge of interest in an area of so-called quantum machine learning~\cite{biamonte2017quantum, arunachalam2023role}, where the concept of quantum learning may innovate a new frontier of knowledge~\cite{chrisley1995quantum} and accelerate corresponding learning task~\cite{huang2022quantum,dong2008quantum}. At the heart of quantum computation is quantum algorithm, which has been extensively developed to tackle a wide range of problems with certain degrees of speedup, such as data fitting~\cite{wiebe2012quantum, wiebe2014quantum}, simulating quantum systems~\cite{low2017optimal, low2019hamiltonian, berry2007efficient, berry2012black, berry2014high},  solving linear systems~\cite{harrow2009quantum, childs2017quantum}, solving differential equations~\cite{childs2021high}, etc. 

Recently, there has been growing attention towards the application of quantum computation in topological data analysis~\cite{lloyd2016quantum, schmidhuber2022complexity, ubaru2021quantum, gunn2019review, mcardle2022streamlined, ameneyro2022quantum, crichigno2022clique, cade2021complexity}. The primary motivation is that, the topological structure of given data points are based on the discrete objects, e.g., simplicial complexes, built from a given data points. The number of simplicial complexes can be exponential in the number of data points, which inherently induces a daunting challenge for any computational method. Thus, it is highly motivating to explore the quantum method to solve the corresponding problem, as quantum computers naturally operate on an exponentially vast space (e.g., Hilbert space of qubits). However, the difficulty of computing/estimating Betti numbers (and even normalized Betti numbers) turns out to be more severe. The striking results in~\cite{schmidhuber2022complexity} showed that even estimating Betti numbers up to some multiplicative accuracy is NP-hard. Hence, in the generic case, exponential speedup is not possible under widely believed complexity-theoretic assumption, and it is only possible in specific settings which may not be so natural in the context of TDA~\cite{schmidhuber2022complexity}. Given such hardness of the problem, it is still interesting to explore more strategies to solve them. A thorough review plus detailed analysis for topological data analysis and related algorithms can be found in~\cite{schmidhuber2022complexity, gunn2019review}. 

Recently, the authors in~\cite{nghiem2023improved} introduced a method that allows to estimate the largest eigenvalue of a given matrix $A$ that removed the exponential barrier presented in a previous work of the same authors~\cite{nghiem2022quantum}. As remarked in~\cite{nghiem2023improved}, the new method, which is based on elementary operations with block encoding, is surprisingly powerful. Thus, it is of interest to explore further the capacity of the algorithm via block encoding. Here, we leverage such a method to the context of Betti numbers and normalized Betti numbers estimation. The analysis that will be performed shows that, while this method is slower than~\cite{schmidhuber2022complexity,gunn2019review,lloyd2016quantum}, it can be faster by some polynomial factors (in the number of data points $n$) than the best known classical approach in the cases where the given graph has non-small Betti numbers. Further, the method we introduce here requires considerably lower circuit depth than previously proposed method.

\section{Main Framework}
\subsection{Overview of Prior Works}
In accordance with  \cite{schmidhuber2022complexity}, we consider $n$ data points. Let $\sigma$ denotes the simplicial complex built from $n$ points and $S_k$ denotes the set of $k$-simplex in $\sigma$, given a certain length scale $\epsilon$. Additionally, as in \cite{lloyd2016quantum}, we assume a membership oracle that allows us to query the simplexes, based on the pairwise distance between points. Quantumly, the oracle acts as following:
\begin{align}
    O_f \ket{s} \ket{0} = \ket{s} \ket{0/1} 
\end{align}
where $\ket{0/1}$ indicates if the simplex $\ket{s}$ doesn't belong/ belongs to the filtration. Such oracle allow us to prepare the following state based on Grover method:
\begin{align}
    \ket{S_k} = \frac{1}{\sqrt{|S_k|^{1/2}}} \sum_{s \in S_k} \ket{s}
\end{align}
which is a key recipe in previous works \cite{schmidhuber2022complexity, lloyd2016quantum, gunn2019review}. We remind that from \cite{gunn2019review, schmidhuber2022complexity}, the complexity of a Grover operator is 
\begin{align*}
    \mathcal{O} ( n k )
\end{align*}

To see how the above state is obtained, we note that the oracle allows us to first arrive at the following:
\begin{align}
   \ket{\phi} = \frac{1}{ \sqrt{ C_{k,n}  } } \Big(  \sum_{s \in S_k} \ket{s} \ket{1} + \sum_{s \notin S_k} \ket{s} \ket{0} \Big) 
\end{align}
where 
$$ C_{k,n} = \binom n{k+1} $$

One can see that, if we perform the measurement on the second register in the above state, then we can obtain the state $\ket{S_k}$ with probability $\frac{C_{k,n}}{ S_k } $ which could be amplified to $ \sqrt{ \frac{C_{k,n}}{ S_k }  }$ using Grover method. Therefore, the cost for preparing the state $\ket{S_k}$ is
$$ \mathcal{O} \Big( nk \sqrt{ \frac{C_{k,n}}{ S_k }  }  \Big) $$
From the state $\ket{S_k}$, we use CNOT and ancilla system to prepare:
\begin{align}
    \ket{S_k} = \frac{1}{\sqrt{|S_k|} } \sum_{s \in S_k} \ket{s}_A \ket{s}_B
\end{align}
where the subscript $A, B$ is for convenience purpose, e.g, denoting separate system. Tracing out the second register ($B$ register), we obtain the maximally mixed state in the $A$ register
$$ \rho_k = \frac{1}{|S_k|} \sum_{s \in S_k} \ket{s} \bra{s}_A $$

As done in LGZ algorithm \cite{lloyd2016quantum}, performing quantum phase estimation (with extra ancilla $\ket{\bf 0}$ for storing phase values) with the above state and exponentiated Hodge operator $\exp(i \Delta_k)$, allow us to estimate the normalized $k$-th Betti number, $\beta_k/S_k$ from the sampling of phase register, e.g, the probability of measuring zero is:
\begin{align}
    p_0 = \frac{ \beta_k}{ |S_k |}
\end{align}

\subsection{Alternative Methods for Estimating Betti Numbers}

We remark that, based on \cite{gunn2019review}, such probability could also be find as:
\begin{align}
    p_0 = \Tr\Big( \ket{\bf 0}\bra{\bf 0} \otimes \mathbb{I}_A (   U_{PE} \ket{\bf 0} \bra{\bf 0} \otimes   \rho_k U_{PE}^\dagger )   \Big) 
    \label{eqn: p0}
\end{align}
where $U_{PE}$ denotes phase estimation procedure with $\exp(i \Delta_k)$ (on the register $\ket{\bf 0}$ and the first $A$ register $\ket{s}_A$), and $\mathbb{I}$ is identity operator acting on corresponding systems, $A$ and $B$. 

We remark that the essential step is preparing the state $\ket{S_k}$. As pointed out in \cite{lloyd2016quantum}, \cite{schmidhuber2022complexity}, since it is based on Grover method, such construction is only efficient in the case where $S_k$ is sufficiently large, e.g, clique-dense regime, which certainly restrict the applicability of quantum algorithm originally introduced in \cite{lloyd2016quantum}. Here, we observe that the recently introduced method \cite{nghiem2023improved} allow us to achieve the goal of avoiding such subtleties. Roughly speaking, instead of performing measurement to obtain the desired state that is entangled to $\ket{1}$, we would perform certain operations on the state $\ket{\phi}$ directly and extract the desired information from there. In the above state $\ket{\phi}$, we use extra register to copy the strings from the first register:
\begin{align}
    \ket{\phi} = \frac{1}{ \sqrt{ C_{k,n}  } } \Big(  \sum_{s \in S_k} \ket{s}_A \ket{1}  + \sum_{s \notin S_k} \ket{s}_A \ket{0} \Big) \ket{0}_B  \rightarrow \ket{\phi_1} \frac{1}{ \sqrt{ C_{k,n}  } } \Big(  \sum_{s \in S_k} \ket{s}_A \ket{1} \ket{s}_B + \sum_{s \notin S_k} \ket{s}_A \ket{0} \ket{s}_B \Big) 
    \label{eqn: 7}
\end{align}

Now we consider the same phase estimation unitary acting on the system $A$ (plus corresponding extra phase register $\ket{\bf 0}$), plus trivial action on the remaining system. We obtain the following state:
\begin{align}
    \ket{\phi_2} = \frac{1}{ \sqrt{ C_{k,n}} } \Big(  \sum_{s \in S_k} (U_{PE} \ket{\bf 0}  \ket{s}_A )\ket{1} \ket{s}_B + \sum_{s \notin S_k} (U_{PE} \ket{\bf 0} \ket{s}_A) \ket{0} \ket{s}_B \Big) 
    \label{eqn: 8}
\end{align}
If we trace out the $B$ register, we would obtain the density matrix:
\begin{align}
\label{eqn: rho}
    \rho =  \frac{1}{C_{k,n}} \Big(  \sum_{s \in S_k} U_{PE} \ket{\bf 0} \ket{s}\bra{s}_A \ket{\bf 0} U_{PE}^\dagger \otimes \ket{1}\bra{1} +  \sum_{s \notin S_k} U_{PE} \ket{\bf 0} \ket{s} \bra{s}_A U_{PE}^\dagger \otimes \ket{0}\bra{0} \Big) 
\end{align}
which contains our 'desired' mixed state entangled with $\ket{1}\bra{1}$. The following lemma allows us to block-encode the above density operator, for which the proof could be found in~\cite{gilyen2019quantum} (see their Lemma 45). 
\begin{lemma}[\cite{gilyen2019quantum}]
\label{lemma: improveddme}
Let $\rho = \Tr_A \ket{\Phi}\bra{\Phi}$, where $\ket{\Phi} \in  \mathbb{H}_A \otimes \mathbb{H}_B$, and thus $\rho$ is a density matrix that acts on states  in $ \mathbb{H}_B$. Given a unitary $U$ that generates $\ket{\Phi}$ from $\ket{\bf 0}_A \otimes \ket{\bf 0}_B$, then there exists a procedure that constructs an exact unitary block encoding of $\rho$ in complexity $\mathcal{O}(T_U + \log(n) )$ where $n$ is the dimension of $\mathbb{H}_B$. 
\end{lemma}
The next step is to block encode the operator $\ket{\bf 0}\bra{\bf 0 } \otimes \mathbb{I}_A $, which has been done in \cite{gilyen2019quantum}. In \cite{rall2020quantum}, the author shows the following result:
\begin{lemma}[\cite{rall2020quantum}]
\label{lemma: rall}
    Given the block encoding $U$ of some matrix $A$ (whose norm less than 1) and a unitary $U_\rho$ that satisfies 
    $$  \rho =  \Tr_{M} (U_\rho \ket{0}_M \ket{0}_N  \bra{ 0}_M \bra{0}_N U_\rho^\dagger), $$ 
    then the quantity $ \Tr(A \rho)$ can be estimated up to an additive error $\epsilon$ using a circuit of size 
    $$ \mathcal{O}\Big( \frac{T_U  +T_\rho}{\epsilon}\Big),$$
    where $T_\rho$ is the time required to construct $U_\rho$ and $T_U$ is the time required to construct $U$. 
\end{lemma}
The last recipe we need is the result from \cite{camps2020approximate}, which shows how to block encode tensor product of operators given their unitary block encoding respectively.
\begin{lemma}[\cite{camps2020approximate}]
    Given unitary block encoding $\{U_i\}_{i=1}^m$ of multiple operators $\{M_i\}_{i=1}^m$ (assumed to be exact encoding). Then there is a procedure that produces a unitary block encoding operator of $\bigotimes_{i=1}^m M_i$, which requires a single use of each $\{U_i\}_{i=1}^m$ and $\mathcal{O}(1)$ SWAP gates. 
\end{lemma}
Now we can see clearer that the setting is similar to what has been proposed in \cite{nghiem2023improved}. Given the block encoding of $\ket{\bf 0}\bra{\bf 0} \otimes \mathbb{I}_A $ and of some $2 \times 2$ matrix $M$ that we can choose, we can use Lemma \ref{lemma: rall} to prepare the block encoding of $\ket{\bf 0}\bra{\bf 0} \otimes \mathbb{I}_A \otimes M$. Then Lemma \ref{lemma: rall} can be used to estimate the following quantity:
$$ \Tr\Big( (\ket{\bf 0}\bra{\bf 0} \otimes \mathbb{I}_A \otimes M \cdot \rho )     \Big)  \equiv b $$
Using 
$$  \rho =  \frac{1}{C_{k,n}} \Big(  \sum_{s \in S_k} U_{PE} \ket{\bf 0} \ket{s}\bra{s}_A \ket{\bf 0} U_{PE}^\dagger \otimes \ket{1}\bra{1} +  \sum_{s \notin S_k} U_{PE} \ket{\bf 0} \ket{s} \bra{s}_A U_{PE}^\dagger \otimes \ket{0}\bra{0} \Big)  $$
and that
$$ p_0 = \Tr\Big( \ket{\bf 0}\bra{\bf 0} \otimes \mathbb{I}_A (   U_{PE} \ket{\bf 0} \bra{\bf 0} \otimes   \rho_k U_{PE}^\dagger )   \Big)  $$

We get that:
\begin{align}
    \Tr\Big( (\ket{\bf 0}\bra{\bf 0} \otimes \mathbb{I}_A \otimes M \cdot \rho )     \Big)  &= p_0 \frac{|S_k|}{C_{k,n}} \Tr(M\ket{1}\bra{1}) + \frac{p_1}{C_{k,n}} \Tr(M\ket{0}\bra{0}) \\
    &= \beta_k \frac{\Tr(M \ket{1}\bra{1})}{C_{k,n} } + p_1 \frac{\Tr(M \ket{0}\bra{0})}{C_{k,n} } 
\end{align}
where 
$$ p_1 = \Tr\Big(  \ket{\bf 0}\bra{\bf 0} \otimes \mathbb{I}_A  \cdot  \sum_{s \notin S_k} U_{PE} \ket{\bf 0} \ket{s} \bra{s}_A \bra{\bf 0} U_{PE}^\dagger    \Big) $$
and we use the fact that $p_o  |S_k| = \beta_k$. 

We remark that the value of $b$ heavily depends on the matrix $M$. If we choose two different matrices $M_1$ and $M_2$, then repeating the same procedure for both, we then have:
\begin{align}
    b_1  = \beta_k \frac{\Tr(M_1 \ket{1}\bra{1})}{C_{k,n} } + p_1 \frac{\Tr(M_1 \ket{0}\bra{0})}{C_{k,n} } \\
    b_2 = \beta_k \frac{\Tr(M_2 \ket{1}\bra{1})}{C_{k,n} } + p_1 \frac{\Tr(M_2 \ket{0}\bra{0})}{C_{k,n} }  
\end{align}

Given that we know $M_1, M_2$, the above system form a $2 \times 2$ linear systems, i.e, 
\begin{align}
    \begin{pmatrix}
        \frac{\Tr(M_1 \ket{1}\bra{1})}{C_{k,n} }  & \frac{\Tr(M_1 \ket{0}\bra{0})}{C_{k,n} }  \\
        \frac{\Tr(M_2 \ket{1}\bra{1})}{C_{k,n} } & \frac{\Tr(M_2 \ket{0}\bra{0})}{C_{k,n} } 
    \end{pmatrix} 
    \cdot 
    \begin{pmatrix}
        \beta_k \\
        p_1 
    \end{pmatrix} 
    =
    \begin{pmatrix}
        b_1 \\
        b_2
    \end{pmatrix} 
    \equiv Y_1
    \label{eqn: 14}
\end{align}
which could be solved classically to find $\beta_k$ and $p_1$. We remark one thing that the system is not ideal, as the value of $b_1$ and $b_2$ are estimated via Lemma \ref{lemma: rall}. Therefore, the solution that we find by solving corresponding linear equation are not exact. In order to bound the deviation, we use the following result:
\begin{theorem}[\cite{trefethen2022numerical}]
\label{thm: bound}
    Consider two linear equations $A_1 x_1 = y_1$ and $A_2 x_2 = y_2$. If $A_1$ is non-singular and $|| A_1 - A_2|| \leq 1/||A_1^{-1}||$, then the following holds: 
    \begin{align*}
        \frac{||x_2 - x_1||}{||x_1||} \leq \frac{\kappa}{1-\kappa ||A_2-A_1||/||A_1||} \cdot \Big( \frac{||y_2-y_1||}{|||y_1||} + \frac{ ||A_2-A_1 || }{||A_1 ||}  \Big),
    \end{align*}
    where $||\cdot||$ refers to any norm measure (e.g, $l_1, l_2, l_{\infty}$ ) and $\kappa$ is the conditional number of $A_1$. 
\end{theorem}
Essentially the above theorem quantifies the error deviation of the solution given the corresponding error (given some chosen norm) in matrix $A$ and $b$. In our case, since we know exactly the entries of matrix, e.g, left hand side of Eqn. \ref{eqn: 14}, the only error source is from the right hand side of Eqn. \ref{eqn: 14}. Let $X_1$ denotes the real solution to linear equation \ref{eqn: 14}, and $X_2$ be errornous solution, and $Y_2$ be errornous version of the right hand side of eqn. \ref{eqn: 14}. The above theorem reveals that:
\begin{align}
    ||X_2-X_1|| \leq ||X_1|| \cdot  \kappa \cdot \frac{||Y_2-Y_1||}{|||Y_1||}
\end{align}
Since 
\begin{align}
    A X_1  = Y_1
\end{align}
Using a well-known property that $||x y || \leq ||x|| \cdot ||y||$ for arbitrary norm measure, then we have:
\begin{align}
    || A X_1 || = || Y_1 || \leq ||A || \cdot ||X_1 || \rightarrow \frac{||X_1||}{||Y_1||} \leq \frac{1}{||A||}
\end{align}
So we have:
\begin{align}
    || X_2 - X_1 || \leq \frac{\kappa}{||A||} ||Y_2 - Y_1|| 
\end{align}
The conditional number $\kappa = ||A||\cdot ||A^{-1}||$, so eventually we yield:
\begin{align}
    ||X_2 - X_1|| \leq ||A^{-1} || \cdot ||Y_2 - Y_1 ||
\end{align}
Since each of the value $b_1$ and $b_2$ are estimated up to additive accuracy $\delta$ (in $\l_2$ norm), we have that (using $l_2$ norm)
\begin{align}
    ||Y_2 - Y_1|| \leq \sqrt{2} \delta
\end{align}
The solution $X_2$ contains the approximated solution, which has $\Tilde{\beta_k }$, we then have:
\begin{align}
    || \Tilde{\beta_k} - \beta_k || \leq ||X_2 - X_1|| \leq \sqrt{2} \delta ||A||^{-1}
\end{align}
We want to achieve the multiplicative error of $\epsilon$ in estimating $\beta_k$, then we require $|| \Tilde{\beta_k} - \beta_k || \leq \epsilon \beta_k $. So we need to set:
$$ \delta = \frac{\epsilon \beta_k}{ \sqrt{2} ||A||^{-1} } $$
According to Lemma \ref{lemma: rall}, the scaling resource is $\mathcal{O}(1/\delta)$ which contains factor $||A||^{-1}$. We remind that the matrix $A$:
\begin{align}
    A = 
     \begin{pmatrix}
        \frac{\Tr(M_1 \ket{1}\bra{1})}{C_{k,n} }  & \frac{\Tr(M_1 \ket{0}\bra{0})}{C_{k,n} }  \\
        \frac{\Tr(M_2 \ket{1}\bra{1})}{C_{k,n} } & \frac{\Tr(M_2 \ket{0}\bra{0})}{C_{k,n} } 
    \end{pmatrix} 
\end{align}
Since the entries of $A$ depends on the matrix $M_1, M_2$ that we choose, the value $||A||^{-1}$ is usually not known. A simple way to find out is by noting that the norm $||A||^{-1}$ is one over the smallest singular value of $A$, and singular values of $A$ could be found using classical method. However, we observe that there is a factor $C_{k,n}$, which means that the value of $||A||^{-1}$ is as much as $\mathcal{O}(C_{k,n})$. The following theorem summarizes our result:
\begin{theorem}
\label{thm: thm}
    Given quantum access to pairwise distance between arbitrary two points (at a given length scale $d$) in the $n$ data points in a similar manner with LGZ algorithm \cite{lloyd2016quantum}, the $k$-th Betti number of corresponding simplicial complex $\sigma$ could be estimated up to multiplicative accuracy $\epsilon$ in time complexity
    $$  \mathcal{O}\Big( (nk + \kappa n) \frac{C_{k,n}}{\epsilon \beta_k}   \Big)  $$
    where 
    $$ C_{k,n} = \binom n{k+1} $$
    and $\kappa$ is the conditional number of correspopnding Hodge-Laplacian operator. 
\end{theorem}
The running time for quantum algorithm presented in \cite{schmidhuber2022complexity} is 
$$ \mathcal{O}\Big( \frac{1}{\epsilon} (n^2 \sqrt{ \frac{C_{k,n}}{\beta_k} } + n\kappa \sqrt{ \frac{|S_k|}{\beta_k} })   \Big)   $$
Our method achieves a quadratically slower time in the factor $C_{k,n}/\beta_k$, comparing to \cite{schmidhuber2022complexity}. As mentioned in \cite{schmidhuber2022complexity}, the best classical algorithm to compute $k$-th Betti number has running time:
$$ \mathcal{O} ( C_{k,n}  ) = \mathcal{O}(  \binom n{k+1} ) $$
As mentioned in \cite{schmidhuber2022complexity}, there are several graphs that has $\beta_k$ grows as much as $\mathcal{O}(n^k)$, which means that our method could be considerably faster than classical approach. \\

A very interesting consequence of the above method is that, once we solve the desired linear equation, we can also find the $k$-th Betti number of the complement graph. To see this, we note the following quantity:
\begin{align}
    p_1 = \Tr\Big(  \ket{\bf 0}\bra{\bf 0} \otimes \mathbb{I}_A  \cdot  \sum_{s \notin S_k} U_{PE} \ket{\bf 0} \ket{s} \bra{s}_A \bra{\bf 0} U_{PE}^\dagger    \Big) 
\end{align}
has exactly the same form as $p_0$ (Eqn. \ref{eqn: p0}), which is the normalized Betti number. It is worth noting that, from Eqn. (\ref{eqn: 7}) and Eqn. (\ref{eqn: 8}), the computational state $\ket{s}_A$ is separated into two parts, entangled with $\ket{0}$ and $\ket{1}$ which indicates whether or not $\ket{s}_A$ lies in the complex $\sigma$ at length scale $d$. Therefore, the states $\ket{s}_A$ that are entangled with $\ket{0}$ are those simplices not lying in the complex $\sigma$, which means that they constitute the complement graph to the graph of $\sigma$. Therefore, the meaning of $p_1$ is literally the $k$-th Betti number of such graph.

\subsection{Estimating Normalized Betti Numbers}
The strategy from the previous section could be trivially extended to estimating normalized Betti numbers, which is $\beta_k/|S_k|$. Recall that from equation \ref{eqn: 14}, we have a linear system: 
\begin{align}
    \begin{pmatrix}
        \frac{\Tr(M_1 \ket{1}\bra{1})}{C_{k,n} }  & \frac{\Tr(M_1 \ket{0}\bra{0})}{C_{k,n} }  \\
        \frac{\Tr(M_2 \ket{1}\bra{1})}{C_{k,n} } & \frac{\Tr(M_2 \ket{0}\bra{0})}{C_{k,n} } 
    \end{pmatrix} 
    \cdot 
    \begin{pmatrix}
        \beta_k \\
        p_1 
    \end{pmatrix} 
    =
    \begin{pmatrix}
        b_1 \\
        b_2
    \end{pmatrix} 
    \equiv Y_1
\end{align}
which is mathematically equivalent to:
\begin{align}
    \begin{pmatrix}
        \Tr(M_1 \ket{1}\bra{1} ) & \Tr(M_1 \ket{0}\bra{0}) \\
        \Tr( M_2\ket{1}\bra{1}) & \Tr( M_2 \ket{0}\bra{0}) 
    \end{pmatrix}
    \cdot 
    \begin{pmatrix}
        \beta_k/C_{k,n} \\
        p_1/C_{k,n}
    \end{pmatrix}
    =
    \begin{pmatrix}
        b_1 \\
        b_2
    \end{pmatrix} 
\end{align}
The only difference is that we treat $\beta_k/C_{k,n}$ is a variable instead. From the previous section, we can estimate the value of solution, e.g, $\beta_k/C_{k,n}$, up to additive error $\epsilon$ in time $\mathcal{O}(1/\epsilon)$. In order to find the normalized Betti number, we simply need to multiply $\beta_k/C_{k,n}$ with $C_{k,n}/|S_k|$. In order to estimate $\beta_k/|S_k|$ up to desired accuracy $\delta$, we need to choose $\epsilon = \delta \ (|S_k|/C_{k,n})$. Therefore, we have the main result as following:
\begin{theorem}[Estimating k-th Normalized Betti number]
    Given the same condition as in \ref{thm: thm}, the k-th normalized Betti number $\beta_k/|S_k$ could be estimated up to additive accuracy $\delta$ in time
    $$ \mathcal{O}\Big(  (nk + \kappa n) \frac{C_{k,n}}{\delta |S_k|}   \Big) $$
\end{theorem}
Recall that the best-to-date algorithm \cite{lloyd2016quantum, schmidhuber2022complexity} achieves the running time:
$$ \mathcal{O}\Big(  ( n^2 \sqrt{ \frac{ C_{k,n} } { |S_k| } } + n\kappa ) \frac{1}{\delta} \Big)  $$
for the same task of estimating normalized Betti numbers. In general treatment, our method is quadratically slower. However, in the regime where $|S_k|\approx C_{k,n}$, both method achieves similar running time.  \\

\noindent
\textbf{Discussion:} In general, while both method introduced in the previous two sections achieves slower running time comparing to \cite{schmidhuber2022complexity}, we remark that our algorithm require a quantum circuit of depth $\mathcal{O}( nk + n\kappa )$, which is considerably more efficient than the depth required by \cite{schmidhuber2022complexity, lloyd2016quantum, gunn2019review, ubaru2021quantum}, which is $\mathcal{O}(  n^2 \sqrt{ \frac{ C_{k,n} } { |S_k| } } + n\kappa   )$. Thus, the slower running time of our method could be regarded as a trade-off for circuit depth, which might be more relevant to near-term prospect. In the case of normalized Betti numbers, as mentioned, our method still requires lower depth than known approach, but having slower running time. Howver, in the regime of dense simplices, e.g, $|S_k| \rightarrow C_{k,n}$, the fraction $|S_k|/C_{k,n}$ approaches 1, which means that our method can achieve the same running time as well as circuit depth as known approach, e.g, \cite{schmidhuber2022complexity}.

\section{Acknowledgements}
The author thanks Sam Gunn for meaningful discussion. This work was supported by the U.S. Department of Energy, Office of Science, Advanced
Scientific Computing Research under Award Number DE-SC-0012704. We also acknowledge the support by a Seed Grant from
Stony Brook University’s Office of the Vice President for Research and by the Center for Distributed Quantum Processing.

\bibliography{ref.bib}{}
\bibliographystyle{unsrt}

\end{document}